\documentclass[letterpaper, 10 pt, conference]{ieeeconf}  % Comment this line out if you need a4paper

\IEEEoverridecommandlockouts                              % This command is only needed if 
\usepackage{graphics} % for pdf, bitmapped graphics files
\usepackage{epsfig} % for postscript graphics files
\usepackage{mathptmx} % assumes new font selection scheme installed
\usepackage{times} % assumes new font selection scheme installed
\usepackage{amsmath} % assumes amsmath package installed
\usepackage{amssymb}  % assumes amsmath package installed
\usepackage{color}
\usepackage{subfig}
\usepackage{cite}
\usepackage{graphicx}
\usepackage{siunitx}
\usepackage{soul}

\graphicspath{{figs/}}

\title{\LARGE \bf
Real-Time Spatio-Temporal LiDAR Point Cloud Compression}

\author{Yu Feng$^{1}$, Shaoshan Liu$^{2}$, and Yuhao Zhu$^{1}$% <-this % stops a space
\thanks{$^{1}$These authors are with the Department of Computer Science, University of Rochester, Rochester, NY, USA. {\tt\small yfeng28@ur.rochester.edu}, {\tt\small yzhu@rochester.edu}}%
\thanks{$^{2}$This author is with PerceptIn, Inc., Santa Clara, CA, USA. {\tt\small shaoshan.liu@perceptin.io}}
}

\begin{document}

\maketitle
\thispagestyle{empty}
\pagestyle{empty}

%%%%%%%%%%%%%%%%%%%%%%%%%%%%%%%%%%%%%%%%%%%%%%%%%%%%%%%%%%%%%%%%%%%%%%%%%%%%%%%%

%!TEX root=paper.tex

\newcommand{\website}[1]{{\tt #1}}
\newcommand{\program}[1]{{\tt #1}}
\newcommand{\benchmark}[1]{{\it #1}}
\newcommand{\fixme}[1]{{\textcolor{red}{\textit{#1}}}}

\newcommand*\circled[2]{\tikz[baseline=(char.base)]{
            \node[shape=circle,fill=black,inner sep=1pt] (char) {\textcolor{#1}{{\footnotesize #2}}};}}

\ifx\figurename\undefined \def\figurename{Figure}\fi
\renewcommand{\figurename}{Fig.}
\renewcommand{\paragraph}[1]{\textbf{#1}~~}
\newcommand{\figline}{{\vspace*{.05in}\hline}}

\newcommand{\Alg}[1]{Alg.~\ref{#1}}
\newcommand{\Sect}[1]{Sec.~\ref{#1}}
\newcommand{\Fig}[1]{Fig.~\ref{#1}}
\newcommand{\Tbl}[1]{Tbl.~\ref{#1}}
\newcommand{\Equ}[1]{Equ.~\ref{#1}}
\newcommand{\Apx}[1]{Appendix~\ref{#1}}

\newcommand{\specialcell}[2][c]{\begin{tabular}[#1]{@{}l@{}}#2\end{tabular}}
\newcommand{\note}[1]{\textcolor{red}{#1}}

\newcommand{\greenweb}{{\fontfamily{cmtt}\selectfont GreenWeb}\xspace}
\newcommand{\autogreen}{\textsc{AutoGreen}\xspace}
\newcommand{\proj}{\textsc{ASV}\xspace}
\newcommand{\mode}[1]{\underline{\textsc{#1}}\xspace}

\newcommand{\floor}[1]{\left\lfloor #1 \right\rfloor}
\newcommand{\ceil}[1]{\left\lceil #1 \right\rceil}

\newcommand{\RNum}[1]{\uppercase\expandafter{\romannumeral #1\relax}}

% checkmark and xmark in the pifont package
%\newcommand{\cmark}{\ding{51}}
%\newcommand{\xmark}{\ding{55}}

\begin{abstract}

Compressing massive LiDAR point clouds in real-time is critical to autonomous machines such as drones and self-driving cars. While most of the recent prior work has focused on compressing individual point cloud frames, this paper proposes a novel system that effectively compresses a sequence of point clouds. The idea to exploit both the spatial and temporal redundancies in a sequence of point cloud frames. We first identify a key frame in a point cloud sequence and spatially encode the key frame by iterative plane fitting. We then exploit the fact that consecutive point clouds have large overlaps in the physical space, and thus spatially encoded data can be (re-)used to encode the temporal stream. Temporal encoding by reusing spatial encoding data not only improves the compression rate, but also avoids redundant computations, which significantly improves the compression speed. Experiments show that our compression system achieves 40$\times$ to 90$\times$ compression rate, significantly higher than the MPEG's LiDAR point cloud compression standard, while retaining high end-to-end application accuracies. Meanwhile, our compression system has a compression speed that matches the point cloud generation rate by today LiDARs and out-performs existing compression systems, enabling real-time point cloud transmission.

\end{abstract}

%%%%%%%%%%%%%%%%%%%%%%%%%%%%%%%%%%%%%%%%%%%%%%%%%%%%%%%%%%%%%%%%%%%%%%%%%%%%%%%%
% main paper

\section{Introduction}

LiDAR has become an essential sensor in autonomous machines such as self-driving vehicles, autonomous drones, and robots. LiDARs generate massive amounts of point cloud data. For instance, the Velodyne HDL64E LiDAR generates hundreds of thousands of points each frame, amounting to up to 26 MB of raw data per second. Effectively compressing point cloud in real-time enables autonomous machines to be closely connected with each other and with the cloud, ushering in a new era in distributed and cloud robotics. For instance, efficient point cloud compression would enable offloading compute-intensive perception tasks (e.g., object detection) to the cloud to reduce the perception latency; similarly, collaborative decision making across robots relies on efficient point cloud compression to exchange information.

While prior work mostly focuses on the compression rate~\cite{houshiar20153d, tu2016, gpcc2019}, our work aims to simultaneously improve the compression rate and compression speed while maintaining high accuracy for end-to-end applications of interest (e.g., registration and object detection). High compression speed let the compressed point cloud be transmitted in real-time without local buffering, easing the storage pressure.
%High accuracy let autonomous machines compute accurately from compressed point cloud, which is what ultimately matters from an application's perspective.

%Today's point cloud compression techniques are unsatisfactory because they do not directly target the how point clouds will be used in real applications such as registration and object detection. Instead, they use accuracy metrics such as PSNR that does not directly indicate the application-level accuracy.

We propose a real-time Spatio-temporal point cloud compression technique that delivers high compression rate, maintains high application-level accuracy while delivering a compression speed ($>$10 Hz) that matches/exceeds the LiDAR point cloud generation speed. We use range image~\cite{tu2016} as the basic data representation for point clouds. Range images not only inherently provide a lossless compression of point clouds, but also ``regularize'' the unstructured 3D point cloud data into a structured 2D data structure, which enables computationally-efficient subsequent processing.

Our method exploits the unique redundancies inherent in LiDAR point cloud capturing --- both spatially (within a point cloud) and temporally (across point clouds). We first identify a key point cloud (K-frame) in a point cloud sequence, and transform the rest of the point clouds, which we call predicted clouds (P-frames), into K-frame's coordinate system using IMU measurements. We spatially encode the K-frame by exploiting that many points in real-world scenes lie on the same plane and can be encoded using planes.

Building on top of the spatial encoding of the K-frame, we exploit the fact that consecutive point clouds share a great chunk of overlapped areas of the scene. Thus, the same set of planes could be used to encode points across point clouds. Our temporal encoding scheme reuses planes identified in the K-frame to encode overlapped scenes in P-frame. The temporal encoding also compensates for the inaccuracies in IMU measurements, improving the robustness of the method.

We evaluate the proposed method using the KITTI dataset. Our compression method achieves up to 90$\times$ compression rate, significantly out-performing MPEG's LiDAR point cloud compression standard~\cite{mpeg2018, gpcc2019}. Meanwhile, our compression method operates at least 10 Hz, which matches today's LiDAR point cloud generation speed and is higher than prior methods. The high compression speed is achieved both by avoiding redundant computations (e.g., reusing the spatially encoded planes calculated in the K-frame) and by a careful parallel implementation of our algorithm.

Finally, unlike prior work that evaluates quality metrics such as Peak Signal-to-Noise Ratio (PSNR) that are not directly tied to end-to-end application-level accuracy, our compression method directly focuses on application-level accuracy. We show that our compression system on three point cloud applications---registration, object detection, and segmentation---retains the similar accuracy as the original point cloud while out-performing the accuracies of existing point cloud compression schemes. Our method delivers high application-level accuracy because spatial and temporal encoding inherently preserve the geometry of points in the scene and denoise the point clouds.

\begin{figure*}[t]
  \centering
  \includegraphics[width=2\columnwidth]{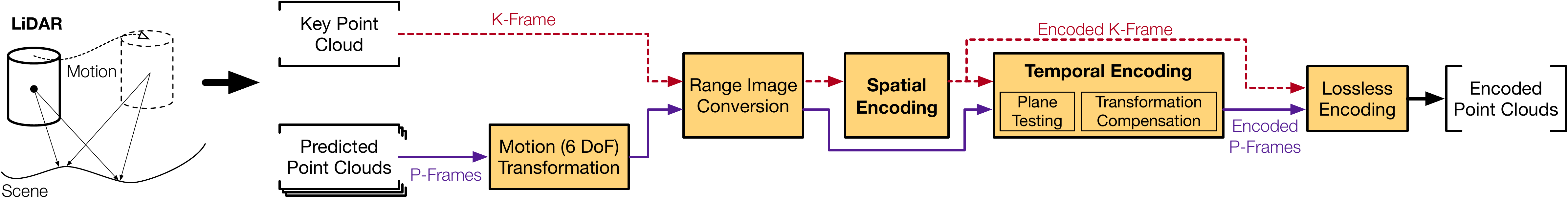}
  %\vspace{-15pt}
  \caption{Overview of our compression system, which compresses a sequence of consecutive point clouds. All the points clouds are converted to range images to accelerate the compression speed. We first spatially encode the key point-cloud (K-frame) in the sequence, typically the middle one. The spatial encoding results of the K-frame are then used to temporally encode the rest of the point clouds, which we call predicted point clouds (P-frames).}
  \label{fig:overview}
\end{figure*}
 
Our main contributions of this paper are as follows:
\begin{itemize}
    \item To the best of our knowledge, this is the first work that leverages both the spatial and temporal redundancies to compress LiDAR points clouds.
    \item The compression method simultaneously achieves higher compression rate, higher compression speed, and higher application-level accuracy than today's compression methods, including the MPEG's point cloud compression standard.
    %\item we proposed a multiple-origin method that expand the applicability of our compression method from single-frame to large-scale point cloud.
\end{itemize}
\section{Related Work}

%\fixme{broken related work.}
%Much of the prior works on PCC can be categorized into two ideas, first is to represent the spatially scattered points using sparse data structures, such as, Octree or hierarchical map. Another idea is to voxelized or regularize points and make them structured. After either form of transformation, different compression methods will be applied on transformed data structure. The following paragraphs briefly summarize the recent progress on PCC.

\paragraph{Unstructured Point Cloud Encoding} Perhaps the most common way to encode point cloud data is to use space-partitioning trees, among which Octree is the most widely used~\cite{huang2008octree, kammeri2012, smith2012, hornung2013, golla2015, thanou2016, gpcc2019}. The G-PCC method in MPEG's point cloud compression standard falls into this category~\cite{gpcc2019}. Each Octree leaf node could be encoded by either a single occupancy bit, which could be lossless if each leaf node contains exactly one point, or by plane extraction, which preserves more details if each leaf node contains multiple points. G-PCC provides both options. Based on the space-partitioning tree representation, prior work has explored various methods to reduce redundant information, such as 2D projection~\cite{hornung2013} or surface fitting~\cite{smith2012}.

Prior work also exploited temporal redundancies in space-partitioning trees such as XORing the two consecutive Octrees~\cite{kammeri2012}, using motion compensation in 3D space~\cite{thanou2016}, or applying video compression directly~\cite{golla2015}.

Other unstructured point cloud representations include shape adaptive wavelet~\cite{ochotta2004, daribo2011} and hierarchical height map~\cite{morell2014, hornung2013}. While effective in certain use-cases, the downside of unstructured representations is that they do not exploit the unique characteristics exposed by LiDAR point clouds, leading to generally low compression rate.

\paragraph{Structured Point Cloud Compression} Instead of encoding point clouds using space-partitioning trees, another category of compression methods convert point clouds into 2D images using spherical projection~\cite{tu2016, tu2019dnn, sun2019} or orthogonal projection such as the V-PCC method in the MPEG's standard~\cite{vpcc2019, vpcc2020}. Existing image/video compression methods are then used to further compress the projected images~\cite{vpcc2019, tu2016, sun2019}. However, directly applying image/video compression algorithms does not preserve the spatial information inherit in the point cloud, and thus generally results in low application accuracy.
%Instead, our work proposes two key insights to encoding both spatial and temporal redundancies in point cloud and achieve significantly high application accuracy and compression rate.

%\paragraph{Video Compression} Our spatio-temporal compression scheme is inspired by video compression, which leverages spatial redundancies to compress within a frame (i.e., I-frames) and uses temporal redundancies to further improve the compression rate (across P-frames and B-frames)~\cite{??}. However, we find that temporal information serves a different purpose in point cloud compression. Instead of improving compression rate, temporal information mainly helps improve compression \textit{speed} in point cloud compression rather than compression \textit{rate}. \fixme{some explanation}.

%\fixme{this paragraph doesn't belong here.}
%\paragraph{Point Cloud Compression in Real-time Applications} Due to arising interests in autonomous driving, robotic teleoperation and AR/VR applications, there are increasing numbers of research focusing on object detection, classification, registration on point cloud in real time[??]. However, real-time point cloud transfer have not been intensively studied yet, and only a handful of studies have focused on end-to-end application accuracy instead of standard metrics. Therefore, an effective and low-complexity algorithm is needed to bridge this gap. Unlike previous studies, we focus on end application precision and provide an unified compression method for LiDAR-generated point cloud.

\section{Spatio-Temporal Compression}
\label{sec:med}

This section introduces our spatial-temporal LiDAR point cloud compression algorithm. We first present an overview of our compression system (\Sect{sec:med:ov}), followed by the detailed designs of the three key components: range image conversion (\Sect{sec:med:range}), spatial encoding (\Sect{sec:med:spatial}), and temporal encoding (\Sect{sec:med:temporal}). Finally, we discuss our parallel implementation that further improves the compression speed  (\Sect{sec:med:opt}).

\subsection{Main Idea}
\label{sec:med:ov}

%\Fig{fig:framework} shows the framework of our spatiotemporal compression method. In our framework, we first convert points' coordinates in a sequence of point clouds from their own coordinate system to the reference coordinate system using IMU information. After the transformation, each point cloud is projected into a range image using the method described in \Sect{subsec:med:projection} then stack all the range images channel-wise. To compress the range image, we first tile the stacked range image and perform temporal encoding (\Sect{subsec:med:temporal}) to remove the tiles that can possibly fit across channels. After the tiles removed, the remaining range images (called residual range image) will perform spatial encoding individually (\Sect{subsec:med:spatial}). In the end, we perform Hoffman encoding on top of all the encoded data. The decompression process is exactly the reverse process as compression as \Fig{fig:framework} shows.

The idea of our compression system is to exploit redundancies both within a point cloud (spatial) and across point clouds (temporal). Spatially, many surfaces in the real-world are planes (e.g., walls and ground); even non-plane surfaces could be approximated by a set of planes. Temporally, consecutive point clouds share a great chunk of overlapped areas of the scene; thus, the same set of planes could be used to encode points across point clouds. While intuitive, exploiting spatial and temporal redundancies in real-time is challenging due to the irregular/unstructured point cloud and the compute-intensive plane fitting process.

%Therefore, encoding points within and across clouds using planes removes redundancies in point clouds and saves space.

% must address two challenges. First, how should we effectively maximally find the planes to fit points? Second, directly operating on massive unstructured points in the 3D space is hardware unfriendly and time consuming.

\begin{figure}[t]
\centering
%\vspace{-10pt}
\includegraphics[width=\columnwidth]{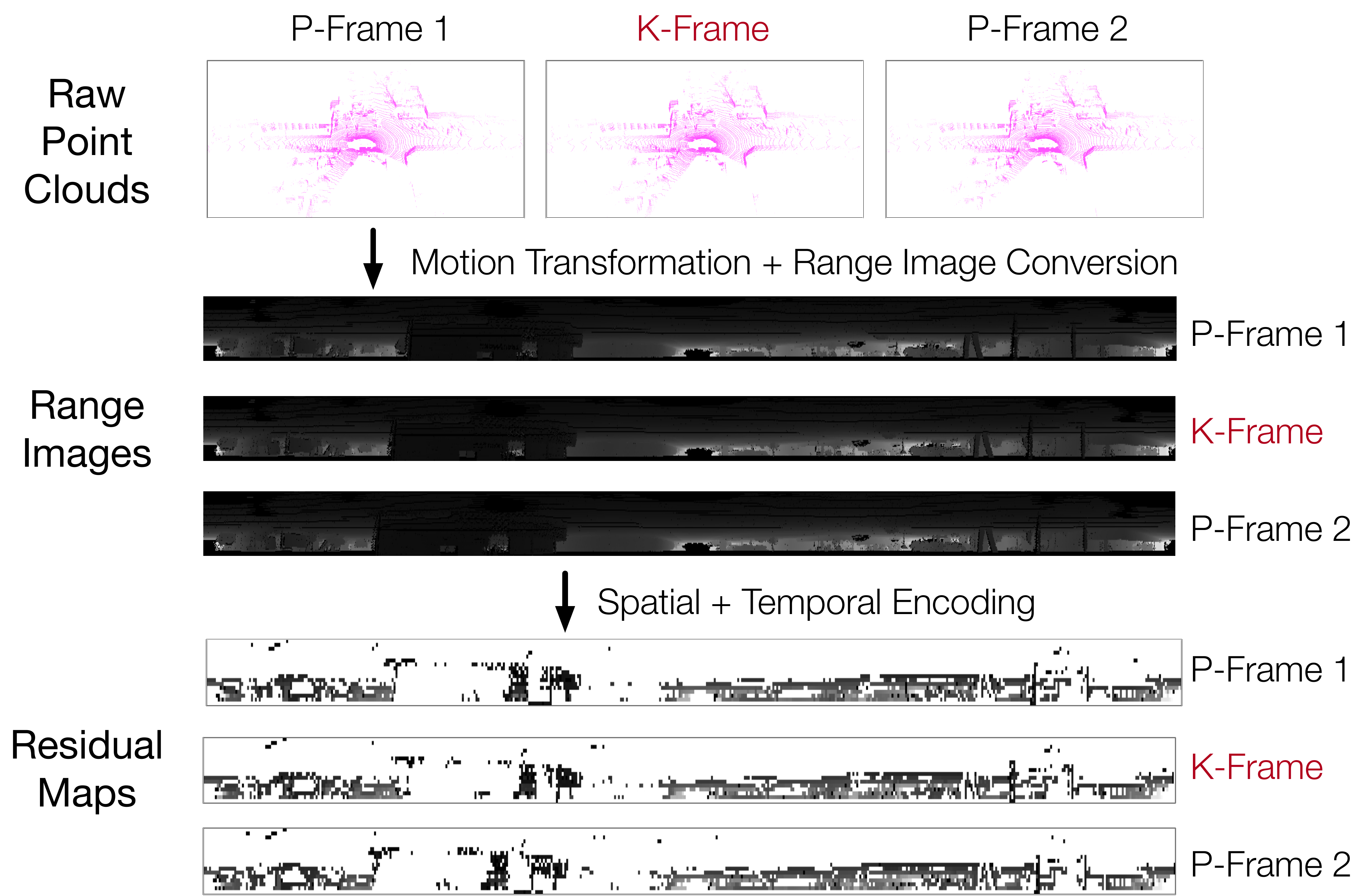}
%\vspace{-10pt}
\caption{Different data structures used in our compression. The raw point clouds are converted to range images. After spatial and temporal encoding, most of the tiles in the range images are plane-encoded; the unfit tiles are left in the residual maps.}
%\vspace{-20pt}
\label{fig:ex}
\end{figure}

We propose a compression system that simultaneously achieves the state-of-the-art compression rate and compression speed while maintaining high application accuracies.~\Fig{fig:overview} provides a high-level overview of our system, which consists of three main blocks: range image conversion, spatial encoding, and temporal encoding.~\Fig{fig:ex} shows the relevant data structures during the encoding process.

% and exploits the spatial redundancy within a point cloud to achieve high compression rate and temporal redundancy across consecutive point clouds to achieve high compression speed.

%Our goal is to compress a sequence of consecutive point clouds captured by a LiDAR.

Given a sequence of consecutive point clouds, we differentiate between two point cloud types: key point cloud (K-frame) and predicted point cloud (P-frame). A sequence has only one K-frame and the rest is P-frames. P-frames are first transformed (both translation and rotation) to K-frame's coordinate system using the IMU measurements. After transformation, each point cloud is converted to a range image~\cite{tu2016} for subsequent computations. The range image not only provides an initial compression to the original point cloud, but also provides a structured representation of the (unstructured) point cloud that is hardware-friendly.

We then spatially encode K-frame by fitting planes; the fitted planes in the K-frame are then (re-)used to temporally encode P-frames, greatly improving the overall compression rate and speed. In order to be robust against transformation errors, which might be introduced due to noisy IMU observations, we propose a set of techniques that compensate the sensor noise and preserve the encoding quality.

In the end, after spatial and temporal encoding, most of the tiles in the range images are plane-encoded; the unfit tiles are left in what we call residual maps (\Fig{fig:ex}). The planes and the residual maps are then further compressed by a lossless compression scheme (e.g., Huffman encoding) to generate the final encoded data.

Overall, in addition to providing high compression rate and speed, our compression system also preserves application accuracy. This is because plane fitting inherently removes noises and outliers in the point clouds without requiring explicitly removing outliers that prior work employs~\cite{sun2019}.

\subsection{Range Image Conversion}
\label{sec:med:range}

We first convert the raw point cloud data to a range image, which essentially converts every point ($x, y, z$) in the 3D Cartesian space to a pixel at coordinates ($\theta, \phi$) in the range image with a pixel value $r$:

\begin{align}
  r=\sqrt{x^2+y^2+z^2};\\
  \theta = arctan(\frac{x}{y})/\theta_{r};~\phi = arccos(\frac{z}{r})/\phi_{r}
\end{align}

\noindent where $\theta_{r}$ and $\phi_{r}$ are the horizontal and vertical resolutions of the LiDAR, respectively.
%\fixme{I don't know if we need to add a offset in each equation, because the range of arctan is [-180, 180] and arccos is [-90,90]}

%For instance, a raw point cloud gathered by HDL-64E LiDAR, which has a \ang{0.08} horizontal resolution and \ang{0.4} vertical resolution as well as a \ang{360} horizontal FOV and a \ang{26.9} vertical FOV, will be converted to an image with a resolution of about \fixme{4500 $\times$ 68}.

A range image naturally compresses the original point cloud, because each point ($x, y, z$) can be encoded with just a range value $r$ of the corresponding pixel in the range image; $\theta$ and $\phi$ are the pixel's coordinates and do not have to be explicitly encoded. If $\theta_{r}$ and $\phi_{r}$ are the same as the resolutions of the LiDAR, range image is a lossless compression of the corresponding point cloud. Mathematically, however, $\theta_{r}$ and $\phi_{r}$ could be any arbitrary positive values; larger $\theta_{r}$ and $\phi_{r}$ would lead to a lower range image resolution, providing a lossy compression of the original point cloud.

In addition to providing an inherent compression scheme, range image brings two key advantages. First, operating on range images is computationally more efficient than directly accessing the point cloud, which requires tree traversals that lead to high cache misses and branch mis-predictions on today's hardware architecture~\cite{xu2019tigris, liu2019point}. Second, adjacent pixels in the range map are likely to lie on the same plane, because they correspond to consecutive scans from the LiDAR. This characteristic allows us to encode the entire range image more efficiently.

%LiDARs sense the distance of surroundings by emitting laser beams and measuring the time of flight response. Beams are emitted in a fixed angular resolution both horizontally and vertically. By knowing the LiDAR's field of view and emission granularity, we can rearrange one point cloud using 2D matrix, in which each row stores the responses received at the same vertical angle and each column stores the responses received at the same horizontal angle, each pixel in the matrix stores the value of sensed distance. For instance, HDL-64E from velodyne can be approximately mapped into a $2000 \times 64$ matrix\cite{velodyne}. However, in order to provide some flexibility into our method, we relax this constrict and allow the matrix to have arbitrary height and width to fulfill different resolution demands. This form of representing the LiDAR point cloud is called range image\cite{tu2016, tu2019dnn, sun2019}. We apply this range image conversion in the first step, which also helps the next two steps in our method.

\subsection{Spatial Encoding}
\label{sec:med:spatial}

The goal of spatial encoding is to encode all the points that lie on the same plane using that plane. Intuitively, many surfaces in the real-world are planes (e.g., walls and ground); non-plane surfaces could be approximated by a set of planes.

In the 3D Cartesian space, a plane can be expressed as:

\begin{equation}
\label{eqn:plane}
   x+ay+bz-c = 0 
\end{equation}

\noindent where ($1, a, b$) is the normal vector of the plane and $\frac{|c|}{\sqrt{1+a^2+b^2}}$ is the distance from the origin (LiDAR center) to the plane. Thus, all the points on the same plane could be encoded with just the three coefficients of the plane. Note that the exact position of each point on the plane is not explicitly encoded. The decoding process would simply have to simulate a ray casting process to find the intersection of a ray and the plane to reconstruct the position of a point.

% As~\Fig{fig:lidar_project} shows, three points, $(a_1, b)$, $(a_2, b)$, and $(a_3, b)$ can be represented by $y = b$ in a 2D space. To decode all the points from a single plane, we can simulate the process of laser projection based on the LiDAR specification and calculate the intersections between beams and plane thus the coordinates of a point cloud. Using this characteristic, we just need to encode which points are in the same plane in the point cloud.

%After range image conversion, we further encode the range image using geometric information and describe pixels using continuous planes. One benefit of using range image is that close points in both horizontal scan and vertical scan are stored adjacently, which provides more opportunities to fit neighboring pixels using one plane. \Fig{fig:encoder} shows the high-level overview of spatial encoding.

To encode the entire point cloud, which contains points that lie in many different planes, we use a ``divide and conquer'' strategy. Specifically, we first uniformly divide the range image into unit tiles (e.g. $4\times4$). We start by fitting a plane for points in the first tile, and gradually grow to include adjacent tiles, essentially forming a bigger tile. Each time we grow, we test whether the plane fit so far can be used to encode all the points in the new (bigger) tile under a predefined threshold. If so, all the points in the new tile are encoded with the plane. Otherwise, we start from the current tile and repeat the process until all the tiles in the range image are processed.

Our spatial encoding process grows tiles horizontally, which we find coalesces many more adjacent tiles than growing vertically. This is inherently because today's LiDARs have a much more fine-grained horizontal resolution than vertical resolution. For instance, Velydone's HDL-64E has a \ang{0.08} horizontal resolution, and a \ang{0.4} vertical resolution. As a result, points in horizontally adjacent tiles are closer to each other and, thus, more likely to fit in the same plane than points from vertically adjacent tiles.

%that can are on the same plane and group them together. From our observation, grouping tiles in both vertical and horizontal dimension complicates the entire algorithm with modest benefit, thus, we only group tiles horizontally.

%The detail procedure is described as following. First, we split a 2D range image into tiles and fit each individual tile into a plane. After calculating the normal for one tile, we test whether the original points can be reconstructed using this plane normal within a defined error threshold. If success, we encode points using the plane coefficients. For two horizontally adjacent tiles, we further test if these two tiles can potentially be described using the same set of plane coefficients, if so, we group adjacent tiles together.

Fitting a plane given points can be naturally formulated as a linear least squares problem~\cite{nievergelt1994total}. While classic iterative methods such as RANSAC~\cite{fischler1981random} are widely used, we find that directly calculating the closed-form solution is generally faster, because deriving the closed-form solution requires less computation and also the computations could be parallelized.

Note that we intentionally do not encode the deltas of plane fitting (i.e., the difference between a true point and a predicted point on the plane). Instead, we find that when a reasonably small threshold is used, discarding deltas effectively \textit{denoises the point cloud}, leading to higher application accuracy than even the original point cloud (\Sect{sec:eva:rate}).

%By considering performance, we choose a time-efficient method to mathematically solve the plan normal instead of iterative fitting methods such as, least-square-fitting. In our method, we first calculate the covariance matrix of the points relative to the centroid value of the points within the tile and then find the smallest Eigenvector of the covariance matrix as the plane normal. In evaluation, our result shows that this method can achieve great runtime performance and parallelism.

%Using the method above, we can encode tiles that can be fitted using plane coefficients, meanwhile, for unfitted tiles, we encode them using median encoding or store their raw values directly. We observe that many horizontal adjacent pixels in a unfitted tile have similar values, we can use a median value to encode the pixels in the same row if the difference between actual pixels and medians with a predefined error threshold. This approach gives us additional compression rate.

\begin{figure}[t]
%\vspace{-8pt}
\centering
\includegraphics[trim=0 0 0 0, clip, width=\columnwidth]{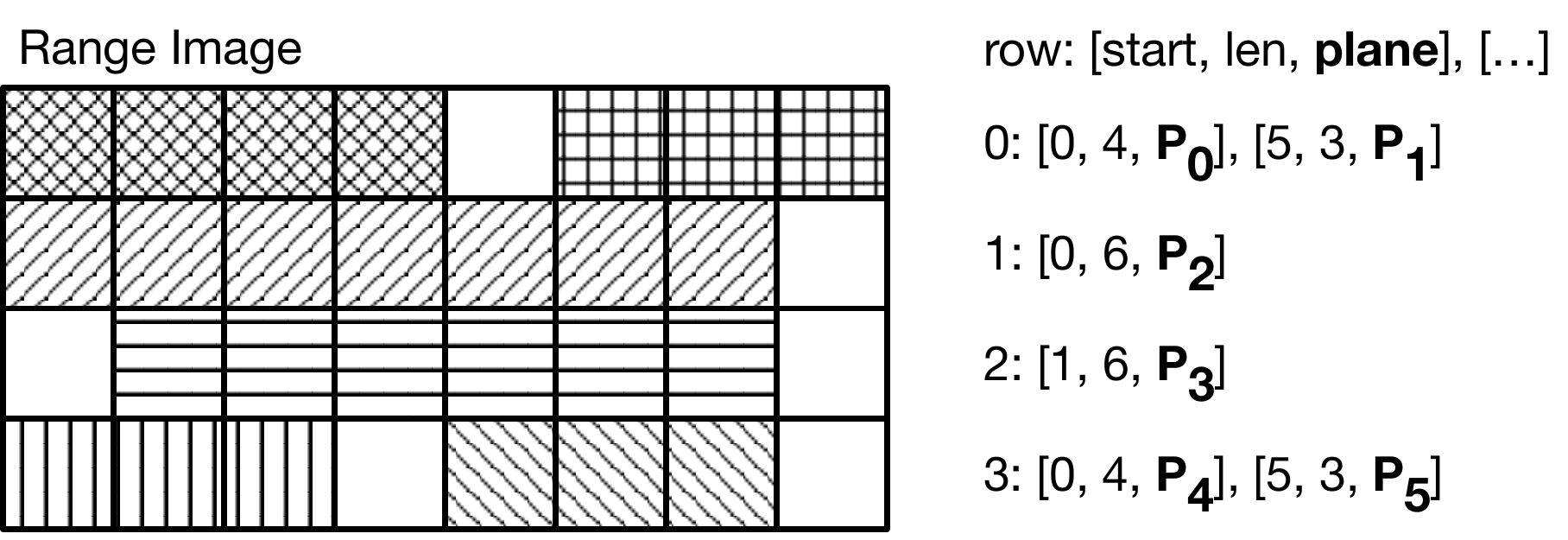}
\caption{A spatial encoding example. The range image on the left is first tiled, and then iteratively planed-fitted. Horizontally adjacent tiles fit by the same plane are shaded by the same stripe pattern. Tiles that are plane-fitted are encoded using the format shown on the right. Points in unfit tiles are encoded individually using their range values (not shown).}
\label{fig:spatial}
%\vspace{-8pt}
\end{figure}

In order to reconstruct/decode the range image later, each row in the range image is encoded with a row ID followed by a set of three-tuples [$s, len, \mathbf{P}$].~\Fig{fig:spatial} provides an example. Each three-tuple corresponds to a sequence of adjacent tiles in that row, starting from $s$ to $s+len$, that are fit by the same plane $\mathbf{P}$, which is parameterized by the three coefficients (\Equ{eqn:plane}). Inevitably, there are tiles that contain points that can not be fit on planes, because, for instance, those points are sparse samples of an irregular surface. These ``unfit'' points are left in what we call a residual map (\Fig{fig:ex}) and are directly encoded using their raw range values.
% or using the base+delta encoding where the base is the median of a tile.

%the spatial encoding module generates two pieces of metadata. First, each fitted plane is associated with a 3-tuple ($s, len, \mathbf{P}$), denoting that points in tiles from $s$ to $s+len$ at row $r$ can fit on the plane. Second, we also generate a tile-map, in which every tile has an element indicating which mode that particular tile is encoded (i.e., fitted on a plane or individually encoded as raw points). \fixme{maybe use a figure here.}
%The tile map along with all the plants and raw points are then losslessly encoded.

%To reconstruct the range image, we need two additional maps, bitmap and tile-map. Bitmap encodes which pixel contains value within the range image, tile-map gives the information what type of technique we used to decode and reconstruct each individual tile. After reconstructing the entire range image, the point cloud can be reconstructed easily by reversing the process of range image conversion.

\subsection{Temporal Encoding} 
\label{sec:med:temporal}

Spatial encoding provides a building block to encode point clouds individually. LiDARs in autonomous machines, however, generate a sequence of point clouds. While it is possible to individually apply spatial encoding to each point cloud, doing so loses opportunities exposed by the temporal correlations across consecutive point clouds.

Consecutive point clouds have large chunks of overlaps, because they are just different samples of the same physical scene. Using the KITTI dataset, we find that on average 99\% of each point cloud is geometrically overlapped with the previous point cloud. Motivated by this observation, temporal encoding encodes a set of consecutive point clouds together. The idea is to use one plane to encode the overlapped scene across multiple consecutive point clouds. Doing so improves both the compression rate and the compression speed by avoiding plane fitting in each point cloud.

%\Fig{fig:stacked_cloud} shows a stacked point cloud from 5 consecutive point clouds in KITTI dataset\cite{geiger2012cvpr}, a great portion of the point cloud is overlapped.

%\begin{figure}[t]
%\centering
%\vspace{-10pt}
%\includegraphics[width=\columnwidth]{misalign}
%\vspace{-10pt}
%\caption{Spatially Mis-aligned Point Clouds}
%\vspace{-20pt}
%\label{fig:misalign}
%\end{figure}

\begin{figure}[t]
\centering
%\vspace{-10pt}
\includegraphics[width=\columnwidth]{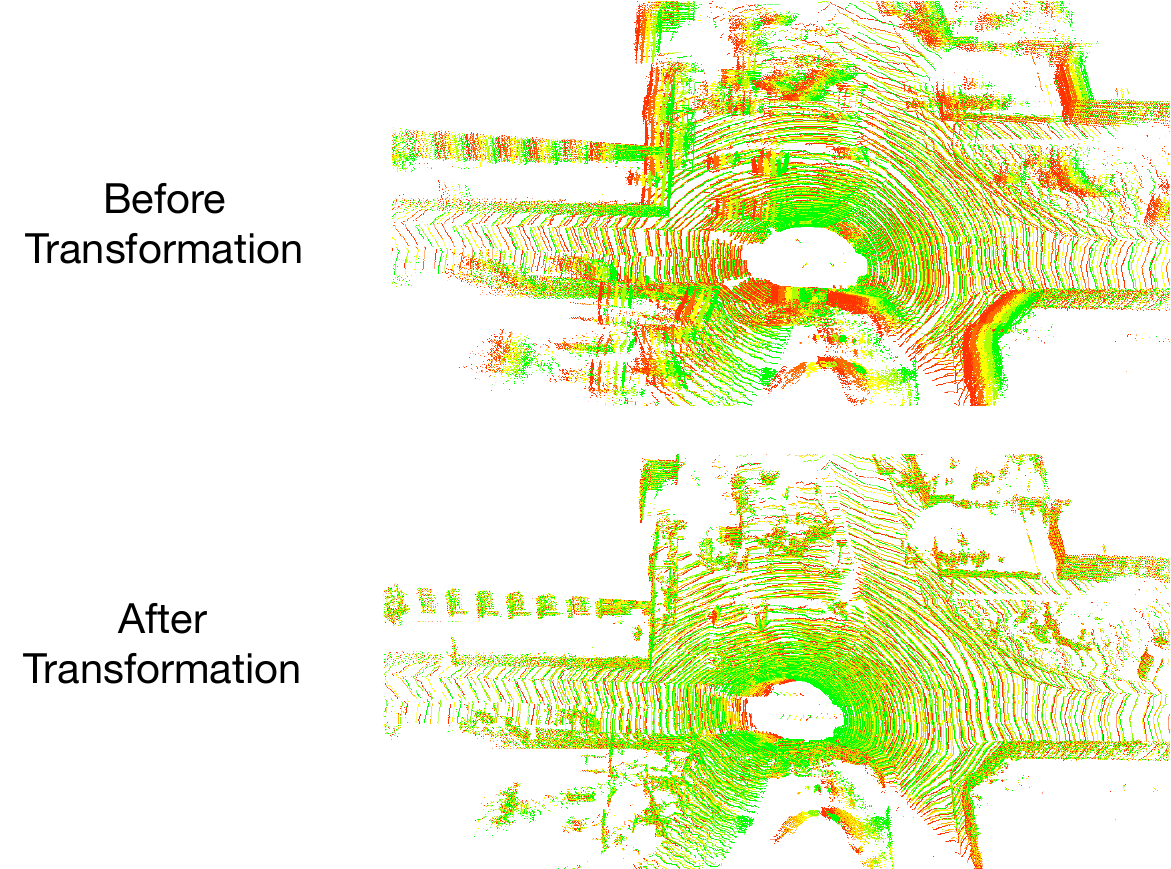}
%\vspace{-10pt}
\caption{Stacking five consecutive point clouds before (top) and after (bottom) motion transformation. Colors indicate different point clouds. Motion transformation better aligns different point clouds in one coordinate system.}
%\vspace{-20pt}
\label{fig:stacked_pcloud}
\end{figure}

\paragraph{Transformation} Each point cloud has its own coordinate system when generated by the LiDAR. In order to fit planes across a sequence of point clouds, we convert all the point clouds to the same coordinate system---by performing a 6 DoF (translation and rotation) transformation.~\Fig{fig:stacked_pcloud} compares the effect of overlaying five consecutive point clouds together in the same coordinate system before and after the transformation. Without motion transformation point clouds at different timestamps are mis-aligned, making temporal encoding challenging.

To unify point clouds in the same coordinate system, we must 1) decide a key point cloud $K$, whose coordinate system is used as the transformation target, and 2) calculate the corresponding transformation matrix $\mathbf{M}_i$ between $K$ and every other point cloud $P_i$.

In our system, we calculate the transformation matrix using the IMU measurements, which provides the translational acceleration ($\mathbf{\hat a}$) and rotational rate ($\mathbf{\hat \omega}$). Using the IMU measurements, we estimate the translation vector $T_{3\times 1}$ as:
\begin{align}
    T_{3\times 1} = \begin{bmatrix}
                    {\overline{\Delta x}} &
                    {\overline{\Delta y}} &
                    {\overline{\Delta z}} 
                    \end{bmatrix}
\end{align}
\noindent where $\overline{\Delta x}$, $\overline{\Delta y}$, and $\overline{\Delta z}$ are translational displacements integrated from $\mathbf{\hat a}$ using the first-order Runge-Kutta numerical method. Similarly, the rotation matrix $R_{3\times 3}$ is estimated as:
%from $T_t$ to $T_{t+1}$, we can form the rotation matrix around each axis and multiply them together.
\begin{align}
    R_{3\times 3} & = \begin{bmatrix}
    cos\large({\overline{\Delta \alpha}}\large) & sin\large({\overline{\Delta \alpha}}\large) & 0\\
    -sin\large({\overline{\Delta \alpha}}\large) & 
    cos\large({\overline{\Delta \alpha}}\large) & 0\\
    0 & 0 & 1
    \end{bmatrix} \nonumber \\ 
    & \times \begin{bmatrix}
    cos\large({\overline{\Delta \beta}}\large) & 0 &  -sin\large({\overline{\Delta \beta}}\large)\\
     0 & 1 & 0 \\
    sin\large({\overline{\Delta \beta}}\large) & 0 &  cos\large({\overline{\Delta \beta}}\large)
    \end{bmatrix} \nonumber \\
    & \times \begin{bmatrix}
    1 & 0 & 0 \\
    0 & cos\large({\overline{\Delta \gamma}}\large) & sin\large({\overline{\Delta \gamma}}\large) \\
    0 & -sin\large({\overline{\Delta \gamma}}\large) & cos\large({\overline{\Delta \gamma}}\large)
    \end{bmatrix}
\end{align}

\noindent where $\overline{\Delta \alpha}$, $\overline{\Delta \beta}$, and $\overline{\Delta \gamma}$ are rotational displacements integrated from $\mathbf{\hat \omega}$ using the first-order Runge-Kutta method.

%At arbitrary timestamp, $T_t$, using the information from IMU, we can obtain the accelerations along three axes, x-, y-, z-axis, ($a^x_t$, $a^y_t$, $a^z_t$) and the angular velocities around these three axes, typically named roll ($v^\alpha_t$), yaw ($v^\beta_t$) and pitch ($v^\gamma_t$). By assuming the constant acceleration at any given time interval, we can estimation the velocity along x-, y-, z-axis, ($a^x_t$, $a^y_t$, $a^z_t$).
%\begin{equation}
%\label{eqn:esti_avg}
%    v^*_{t+1} = v^*_{t} + \frac{( a^*_{t+1} + a^*_t)}{2}\times (T_{t+1} - T_{t})
%\end{equation}
%where $ v^*_{t} $ stands for the linear velocity in one direction, $*$ can be $x$, $y$, $z$.
%To approximate the transformation matrix, $\mathbf{M}$, between two consecutive point clouds, $P_t$ and $P_{t+1}$, we can use first order approximation to estimate the transformation and rotation from $P_t$ to $P_{t+1}$ using \Equ{eqn:esti_avg},
%\begin{equation}
%\label{eqn:esti_avg}
%    \overline{\Delta *}_{(t+1, t)} = \frac{(v^*_{t+1} + v^*_t)}{2}\times (T_{t+1} - T_{t})
%\end{equation}
%where $ \overline{\Delta^*}_{(t+1, t)} $ stands for the rotational or translational difference between two consecutive point clouds, where $*$ can be $x$, $y$, $z$ or $\alpha$, $\beta$, $\gamma$.

We use the middle point cloud in a consecutive point cloud sequence as the key point cloud (K-frame). This minimizes the impact of cumulative IMU sample errors when calculating the transformation matrix. Every other point cloud, which we call predicted cloud (P-frame), is transformed to K-frame's coordinate system by: 

\begin{equation}
    p^{'}_{4\times 1} = \mathbf{M}p_{4\times 1}
    =
    \begin{bmatrix}
    R_{3\times 3}&T_{3\times 1}\\
    0_{1\times 3}&1
    \end{bmatrix}_{4\times 4}p_{4\times 1}
\end{equation}

\noindent where $p_{4\times 1}$ and $p^{'}_{4\times 1}$ denote a point in a predicted cloud before and after transformation, respectively.

%As \Fig{fig:stacked_cloud} shows, consecutive frames contain many redundant information, leveraging this characteristic can effectively increase the compression rate and runtime performance at the same time.

All the $N$ point clouds in a sequence, after transformation, are converted to range images with the same dimension. For the ease of manipulation, we stack the $N$ range images together to form a $N$-channel image.

Note that it is possible that points in a P-frame after transformation could collide, i.e., mapped to the same range image pixel, in which case we preserve the nearest point. On the KITTI dataset, about 4.6\% of the points collide when transforming between two adjacent point clouds, and this percentage increases as the gap two point clouds increases. This suggests that the number of consecutive point clouds that are encoded together ($N$) affects the encoding results. We show the sensitivity to $N$ in~\Sect{subsec:eva:ana}.
%\fixme{I think we need a more theoretical analysis to show that normally, using KITTI data, this is not a huge deal.}

\paragraph{Encoding} We use the same ``divide-and-conquer'' strategy used in spatial encoding to encode across channels (point clouds). A naive implementation would be to fit all the points in a tile across all N channels (e.g., $4 \times 4 \times N$) and then grow to adjacent tiles. However, this approach is susceptible to IMU measurement errors. Inaccurate IMU observations lead to inaccurate point cloud transformations. As a result, points in the same tile across different channels might not end up lying on the same plane, leading to poor plane fitting results.
%~\Fig{fig:misalign} overlays five consecutive point clouds together. The point clouds have been transformed to the same coordinate system. We see misaligned points both in the front of the vehicle and the wall. \fixme{more stuff.}

%Here, we propose our method, which can largely accommodates mis-alignment and achieves high fitting rate while reducing the computation cost.

We propose an effective method to temporally encode across channels while compensating the transformation errors. Our idea is to first fit a tile in the K-frame's channel, and use the fitted plane $Q$ to test against the same tile in each of the other channels. Critically, we hold $Q$'s normal vector constant while varying its distance to the origin (i.e., varying $c$ in~\Equ{eqn:plane}). This effectively compensates the translation error in the IMU measurements. If the relaxed plane $Q'$ (parameterized by $a, b, c'$) fits all the points in a channel under a certain threshold, only $c'$ needs to be encoded for that channel rather than all three plane coefficients.

%If this relaxation is still unable to fit a channel, we perform a fresh plane fitting for that channel and encode the new plane coefficients. This effectively compensates both the translation and rotation errors in transformations.

We apply the same horizontal growing strategy until all the tiles of all the channels in the range image are processed, at which point we remove all the encoded tiles from the range image. The remaining range image $I'$ contains tiles that could not be fitted across channels even after compensating translation errors. We then spatially encode $I'$ channel by channel using the same process described in \Sect{sec:med:spatial}. Effectively, this channel-wise spatial encoding compensates the rotation errors in transformation. In the end, the unfit tiles are left in the residual map (\Fig{fig:ex}), which is further compressed in a lossless fashion along with the fit planes.

Temporal encoding not only provides high compression rate, but also improves the compression speed compared to spatially compressing each point cloud individually. This is because the planes fit in the K-frame are reused in P-frames, reducing the plane fitting overhead.
%We find that about 40\% of the tiles in K-frames can be encoded by reusing previous planes.

%In our method, we only fit the first channel, a.k.a. the points in the first frame, in a tile and allow the plane's offset varies across channels by assuming the plane normal across channels are the same. As \Equ{eqn:plane} shows, if we can reserve the values of $a$ and $b$ and directly plug in the coordinates for the rest of channels, we can calculate the average value $c$ for each channel. After obtaining unique values for each channel, we test whether the coefficients can reconstruct points under the predefined error threshold. If so, all the points in a tile can be expressed using one plane normal, $\{a, b\}$, and $N$ plane offset coefficients, $c$. Previously mentioned technique that merges horizontally adjacent tiles can be applied here as well. By doing so, \fixme{XX\%} of the tiles can be fitted using this technique, which largely reduces the time of fitting. A comprehensive evaluation on performance is shown in \Sect{}.

%\begin{figure}[t]
%%\vspace{-8pt}
%\centering
%\includegraphics[trim=0 0 0 0, clip, width=\columnwidth]{type-category}
%\caption{The percentage of different fitting Types under different compression configurations.}
%\label{fig:obj}
%%\vspace{-8pt}
%\end{figure}

\subsection{Parallel Optimizations}
\label{sec:med:opt}

The speed of the sequential implementation of our algorithm scales linearly with respect to number of channels and angular resolution of the LiDAR. To further improve the compression speed, we exploit the parallelisms exposed by our encoding system and leverage parallel hardware available in modern processors.

At the high level, we exploit both the thread-level parallelism (TLP) and data-level parallelism (DLP). During the range image conversion, we exploit the TLP where each thread is responsible for converting one point cloud into the corresponding range image. During spatial encoding, we leverage TLP where each thread is responsible for encoding a row in the K-frame. During temporal encoding, each thread is responsible for testing planes in a P-frame.

The actual computation in each thread also exposes data-level parallelism such as computing the immediate results (radius, indexes) and the various matrix operations in the plane-fitting and plane-testing processes. Our implementation uses the OpenMP programming model in C++ to exploit both TLP and DLP.

\section{Evaluation Methodology}

\paragraph{Applications and Evaluation Metrics} We evaluate our compression method on three common point cloud applications: registration, object detection, and scene segmentation:
\begin{itemize}
    \item Registration: we use a recent ICP-based registration pipeline~\cite{xu2019tigris} developed using the widely-used PCL~\cite{rusu2011pcl}.
    \item Object Detection: we use VoxelNet~\cite{zhou2018voxelnet}, a Deep Convolution Neural (DNN)-based approach.
    \item Scene Segmentation: we use SqueezeSeg~\cite{wu2018squeezeseg}, a DNN-based approach.
\end{itemize}

We use three evaluation metrics: compression rate over the uncompressed point clouds, compression speed in FPS, and application-level accuracy. We evaluate the application-level accuracy instead of common quality metrics such as PSNR or RMSE because we want to assess how compression affects point cloud applications, which is what ultimately matters.

\paragraph{Dataset} We use the widely-used KITTI dataset~\cite{geiger2012cvpr} for evaluating registration and object detection. We evaluate on all the sequences and frames for comprehensiveness. To evaluate segmentation, we use SemanticKITTI~\cite{behley2019semantickitti}, which augments KITTI dataset for segmentation tasks. We report geometric mean results unless otherwise noted.

%We compare against the two modes that G-PCC provides: one that encodes the point cloud using octree and the other that uses plane fitting to further encode each octree leaf node.
\paragraph{Baseline} We compare against four baselines:
\begin{itemize}
    \item G-PCC: It is a point cloud compression standard proposed by the MPEG~\cite{gpcc2019} specifically designed to compress LiDAR point cloud data. It constructs an Octree for a point cloud and encodes the Octree.
    \item V-PCC: It is a point cloud compression standard proposed by the MPEG~\cite{vpcc2019, vpcc2020} designed to compress dense point clouds used in volumetric rendering. It maps point clouds to images and uses existing video compression to compress the images.
    \item JPEG: It compresses each point cloud's range image individually using the (lossy) JPEG codec~\cite{jpeg2000}.
    \item H.264: It compresses a sequence of point clouds by compressing the corresponding range image sequence using the H.264 video codec~\cite{h264}. We shows results of both the lossy and lossless versions.
\end{itemize}

\paragraph{Variants} Our method can be configured in two modes: the \textit{single-frame} mode that applies only spatial encoding to individual frames and the \textit{streaming} mode that applies both spatial and temporal encoding to a sequence of frames. For both versions, we vary the threshold of plane fitting to form different design points.

\paragraph{Hardware Platform} We implement our compression method in C++ and evaluate the compression speed on both a PC, Intel i5-7500 with 4 cores, and a mobile platform, Nvidia Jetson TX2~\cite{tx2dev}, which represents the compute capability of mobile robots or drones.

%\paragraph{Variants} We evaluate two different variants of our compression method. First variant only compress single individual frames by using spatial encoding and is notated as SF. The second variant compresses a stream of consecutive point clouds, using both spatial and temporal encoding. Without specification, we compress 5 consecutive frames together. Note that, both use Hoffman encoding at the end.
\section{Evaluation}
\label{sec:eval}

We first show the end-to-end accuracy and compression rate of our compression method on three general robotic applications: localization, object detection, and 3D scene segmentation, compared against a range of existing methods (\Sect{sec:eva:rate}). We then demonstrate that our compression speed matches the point cloud generation speed and surpasses other methods (\Sect{subsec:eva:run}). Last, we evaluate the sensitivity of our compression method (\Sect{subsec:eva:ana}).

\begin{figure}[t]
%\vspace{-8pt}
\centering
\includegraphics[trim=0 0 0 0, clip, width=\columnwidth]{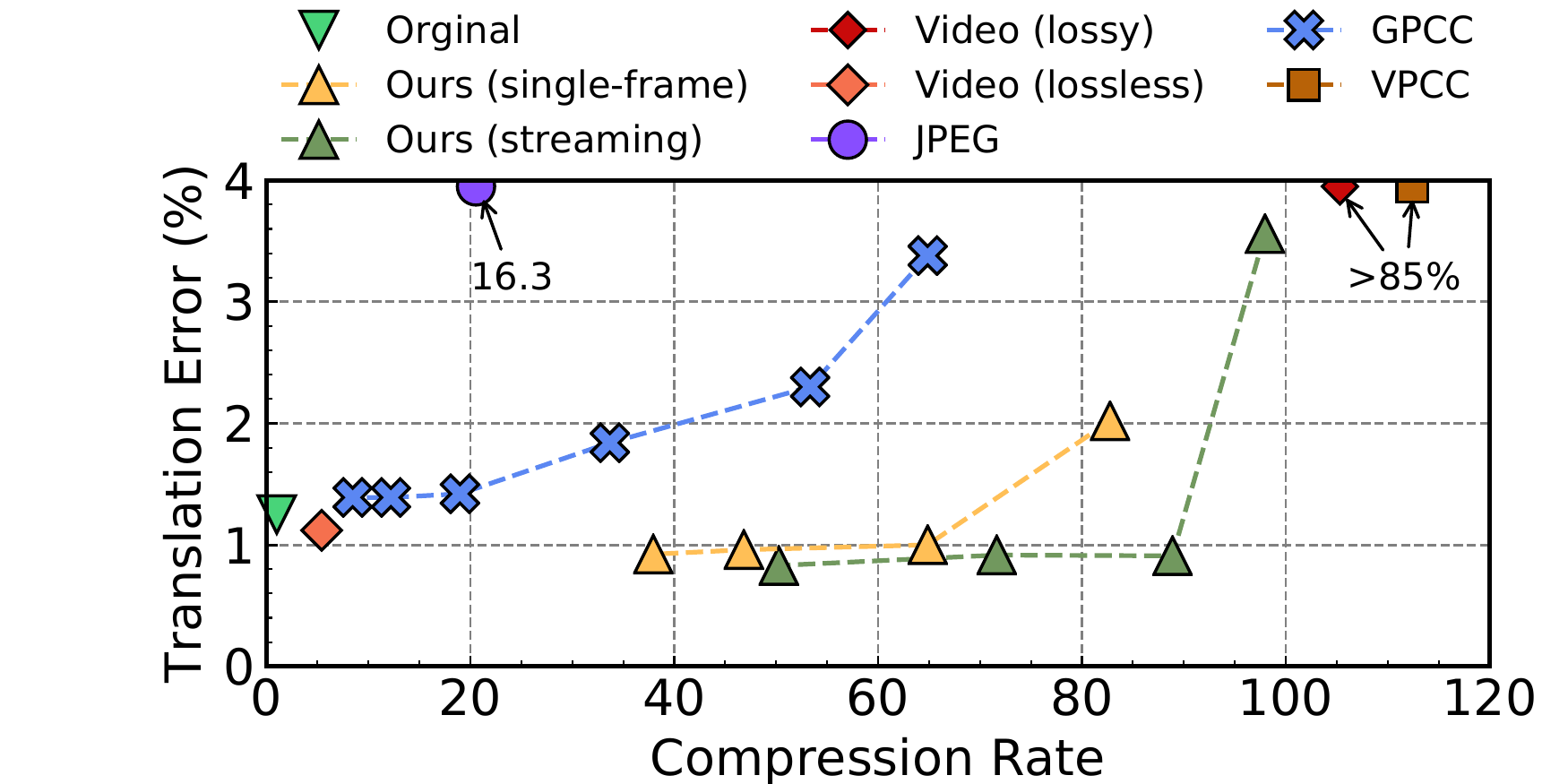}
\caption{Registration translation error and compression rate comparison of various compression methods.}
\label{fig:loc}
%\vspace{-8pt}
\end{figure}

\subsection{Compression Rate vs. Accuracy}
\label{sec:eva:rate}

This section assumes that we compress five consecutive point clouds together unless otherwise noted. We will later study the sensitivity of different frame configurations.

% bitrate of 92 bpp
\paragraph{Localization} Our compression method outperforms other methods in both application accuracy and compression rate.~\Fig{fig:loc} compares the translation error ($y$-axis) against the compression rate ($x$-axis) of different compression methods.

Our method in the streaming mode can achieve an 88.9$\times$ compression rate with only 0.91\% translation error, and the single-frame mode achieves 59.7$\times$ compression rate with 0.96\% translation error. In comparison, the best G-PCC compression has a 1.38\% translation error with only 8.5$\times$ compression rate. Interestingly, our compression methods have lower errors than using the original point clouds (1.25\%). This is because our plane fitting process inherently reduces the noise from the point cloud.

Other baselines including JPEG compression on range images, lossy H.264 video compression, and V-PCC have much higher localization errors ($>$ 16\%) as~\Fig{fig:loc} shows. Although the lossless video compression has better localization accuracy, its compression rate (5.4$\times$) is much lower.
%using the uncompressed data in localization will result in translation error (1.25\%), which is slight highly then using decompressed data by our method. We believe this improvement on localization accuracy is from the side effect of our fitting method, which effectively denoises the original point cloud.

\begin{figure}[t]
%\vspace{-8pt}
\centering
\includegraphics[trim=0 0 0 0, clip, width=\columnwidth]{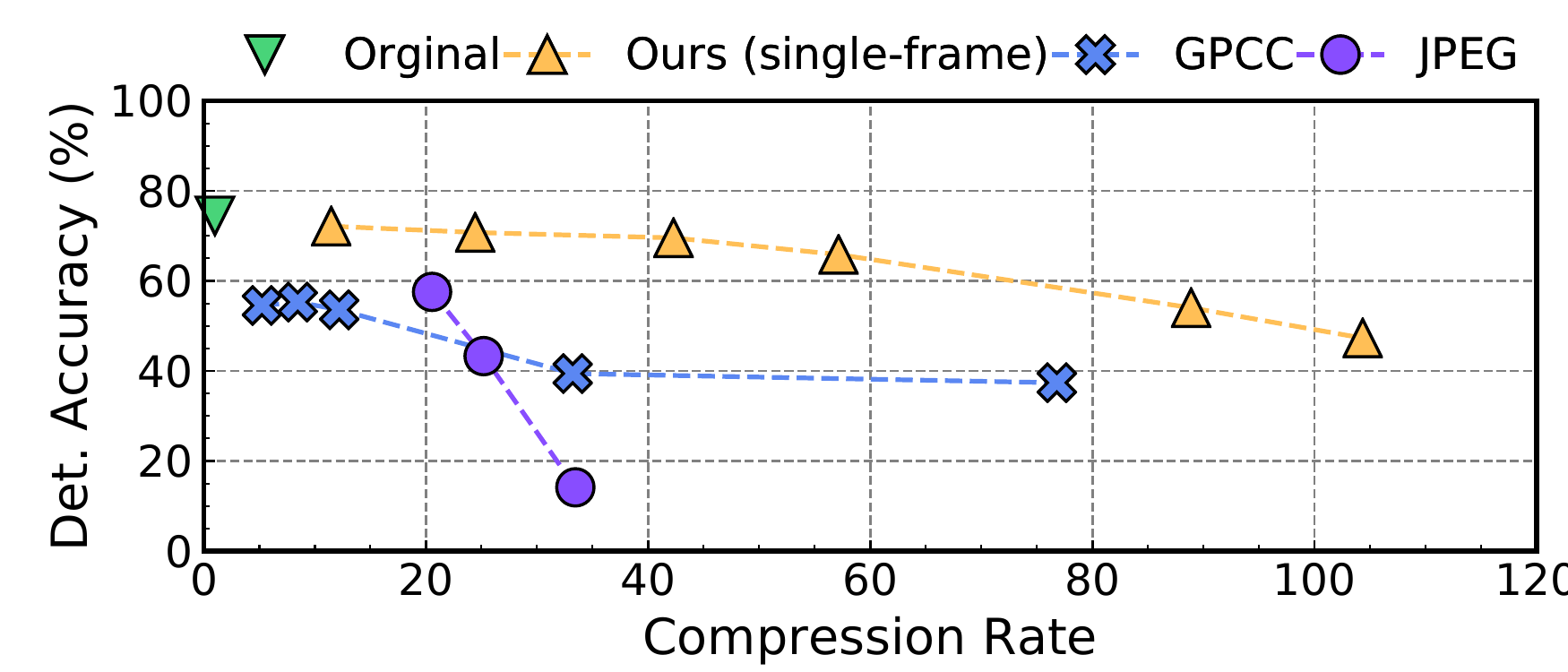}
\caption{The object detection accuracy and compression rate comparison of various compression methods.}
\label{fig:obj}
%\vspace{-8pt}
\end{figure}

\paragraph{Object Detection} On KITTI dataset object detection uses only individual point clouds instead of point cloud sequences. Thus, we present only the single-frame variant of our compression system. For the same reason, V-PCC and H.264 compression methods are not applicable. ~\Fig{fig:obj} compares the object detection accuracy against compression rate across different compression methods. Our method Pareto-dominates the prior methods.

Comparing against the 74.4\% accuracy using the original point clouds, our compression method achieves a comparable accuracy of 72.2\% with a 11.5$\times$ compression rate. In addition, our compression method achieves more than 42.3$\times$ compression rate while still keeping the accuracy over 70\%. In contrast, the best accuracy that G-PCC and JPEG achieve is 42.1\% and 57.6\% with the compression rate of 15.3 and 20.6, respectively.

\begin{figure}[t]
%\vspace{-8pt}
\centering
\includegraphics[trim=0 0 0 0, clip, width=\columnwidth]{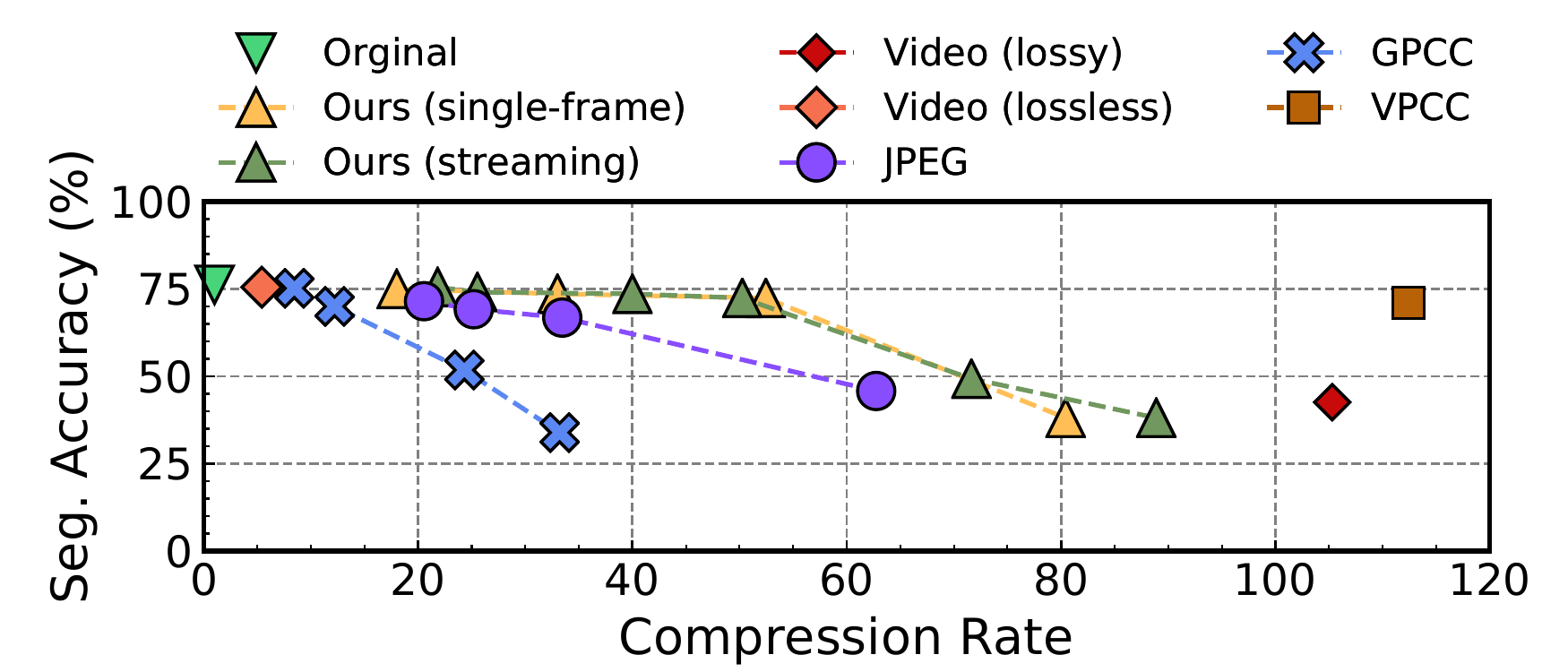}
\caption{The segmentation error and compression rate comparison of various compression methods.}
\label{fig:seg}
%\vspace{-8pt}
\end{figure}

\paragraph{Segmentation}~\Fig{fig:seg} shows the compression rate vs. segmentation accuracy trade-offs across the different compression schemes. Our methods Pareto-dominates other methods except for lossless video compression. In particular, our method achieves a better compression rate (21.8$\times$) than G-PCC (8.5$\times$) and JPEG (20.6$\times$) with a similar accuracy at 75.5\%. The accuracies of G-PCC and JPEG drop quickly as the compression rates increase while our method maintains a high accuracy (72.4\%) even at a compression rate of 50.3$\times$.

Lossless video compression achieves little accuracy drop with only a 5.4$\times$ compression rate; lossy video compression, in contrast, has the highest compression rate---at the expense of over 30\% accuracy drop.

%that our method outperforms others in both compression rate and application accuracy. In particular, our method (75.5\%) and GPCC (75.2\%) achieve an equivalent accuracy as original result (76.2\%), while JPEG has slight lower accuracy (71.5\%). In terms of compression rate, our method achieve better compression rate (21.81$\times$) than GPCC (8.45$\times$) and JPEG (20.55$\times$) with equivalent accuracy. Meanwhile, both our method and JPEG method keep relatively high accuracy (72.4\% and 66.8\%, respectively) at high compression rate of 50.3$\times$ and 33.5$\times$. In comparison, GPCC's accuracy drops quickly after the compression rate increases.

\subsection{Compression Speed}
\label{subsec:eva:run}

\begin{figure}[t]
%\vspace{-8pt}
\centering
\includegraphics[trim=0 0 0 0, clip, width=\columnwidth]{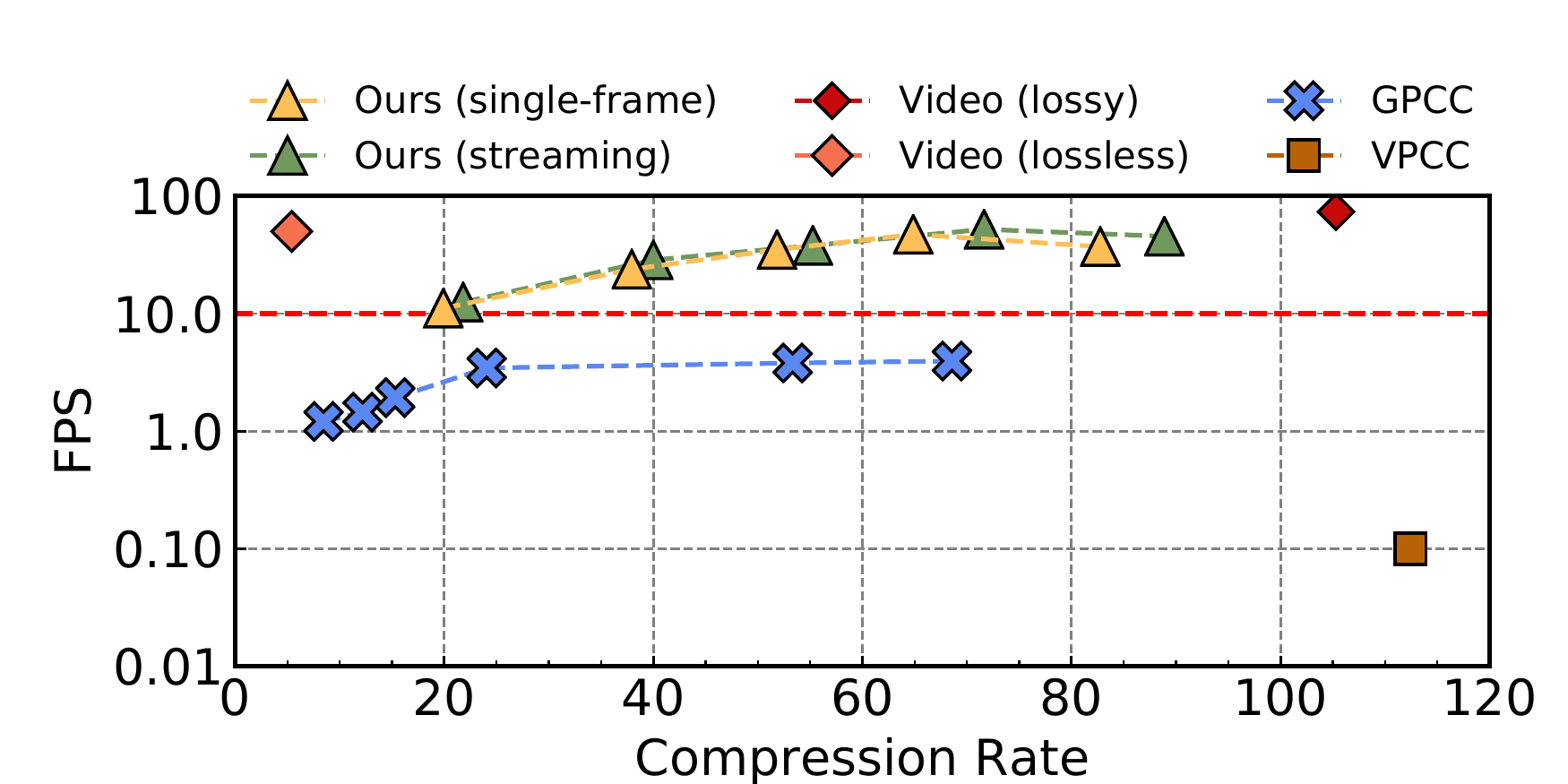}
\caption{Compression speed vs. compression rate of various methods on Intel i5-7500 CPU.}
\label{fig:runtime_desktop}
%\vspace{-8pt}
\end{figure}

\begin{figure}[t]
%\vspace{-8pt}
\centering
\includegraphics[trim=0 0 0 0, clip, width=\columnwidth]{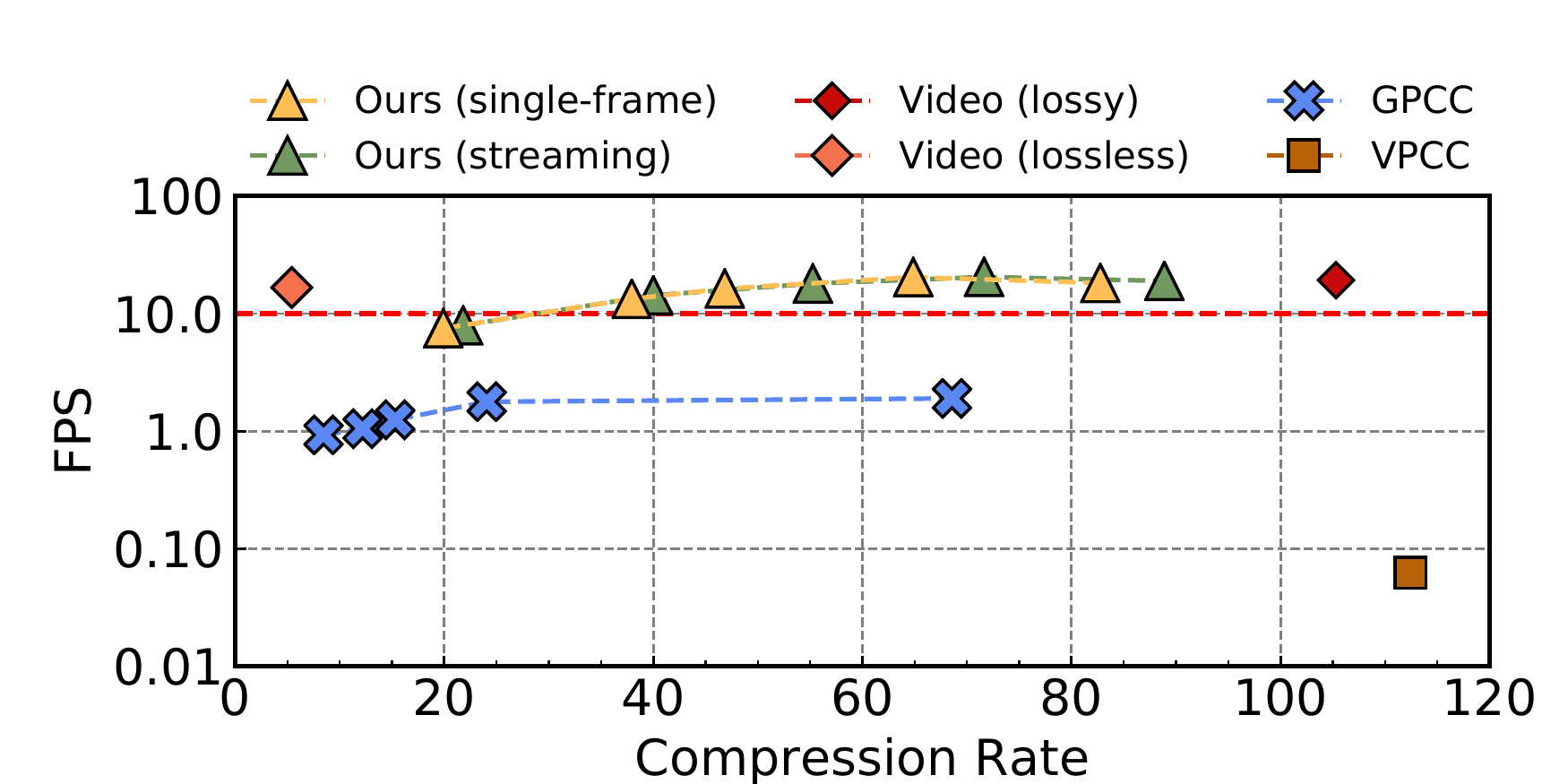}
\caption{Compression speed vs. compression rate of various methods on Nvidia mobile TX2 platform.}
\label{fig:runtime_tx2}
%\vspace{-8pt}
\end{figure}

%To have a fair comparison, all results are implemented in C++ and measured by running compression methods in a single thread. 
\Fig{fig:runtime_desktop} and \Fig{fig:runtime_tx2} show the compression speeds on both a PC and the Nvidia TX2 mobile platform, respectively. Our compression method outperforms G-PCC by about one order of magnitude on both platforms. The compression speed on a PC could be as high as 52.1 FPS, and even on the mobile TX2 the compression speed could be as high as 20.5 FPS. As today's LiDARs generally operate at between 5 Hz to 20 Hz~\cite{velodyne,hdl64e}, our compression method could be executed in real-time as the point clouds are being generated. Lossy and lossless video compressions have a similar compression speed. However, as shown before, they either have a much lower compression rate or lead to much lower application accuracies. V-PCC is much slower than other methods.

%Meanwhile, we demonstrates that our methods can achieve comparable runtime performance comparing to video compression both lossy and lossless compression. However, from previous results,  we show that directly applying video compression is not applicable, due to either low accuracies (lossy video compression) or low compression rate (lossless video compression).

\subsection{Sensitivity Study}
\label{subsec:eva:ana}

\begin{figure}[t]
%\vspace{-8pt}
\centering
\includegraphics[trim=0 0 0 0, clip, width=\columnwidth]{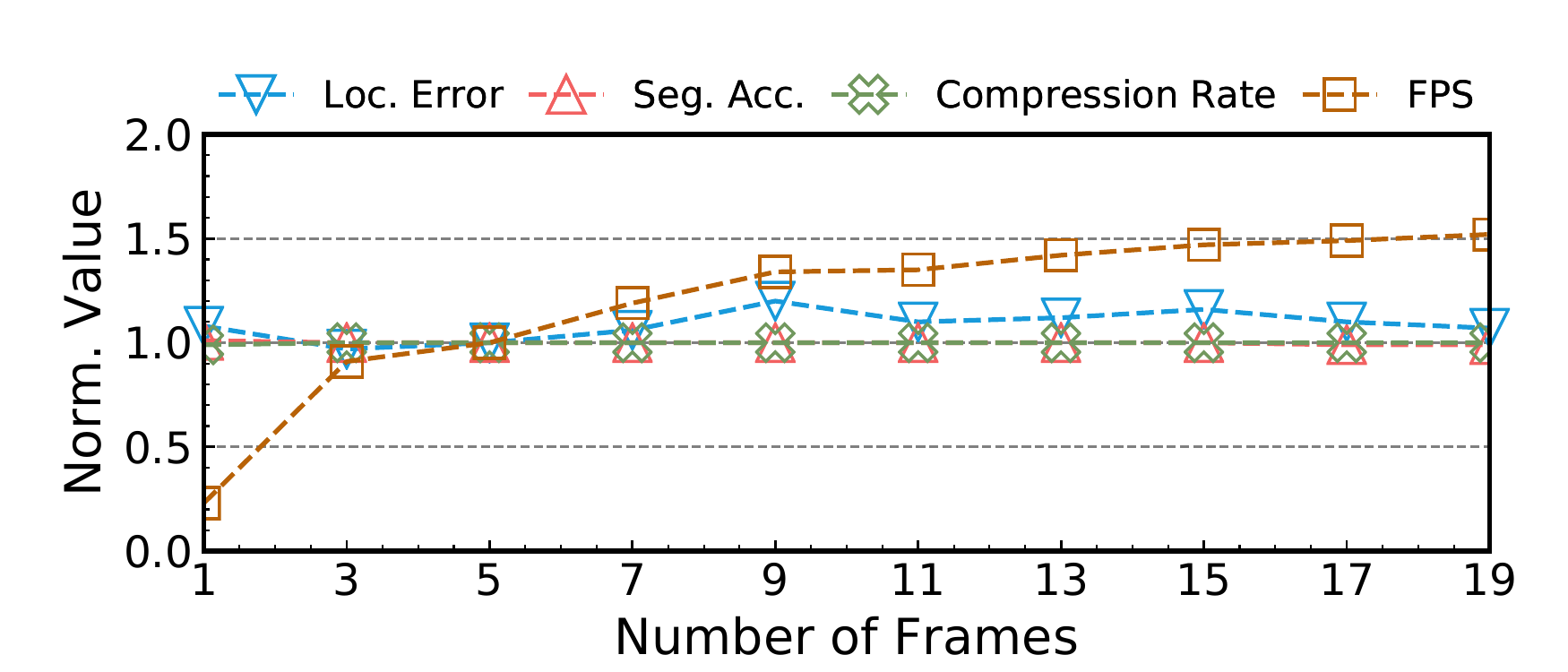}
\caption{Sensitivity study on application accuracy, compression rate, and compression speed by varying the number of consecutive frames that are encoded together.}
\label{fig:sensitivity}
%\vspace{-8pt}
\end{figure}

\begin{figure}[t]
%\vspace{-8pt}
\centering
\includegraphics[trim=0 0 0 0, clip, width=\columnwidth]{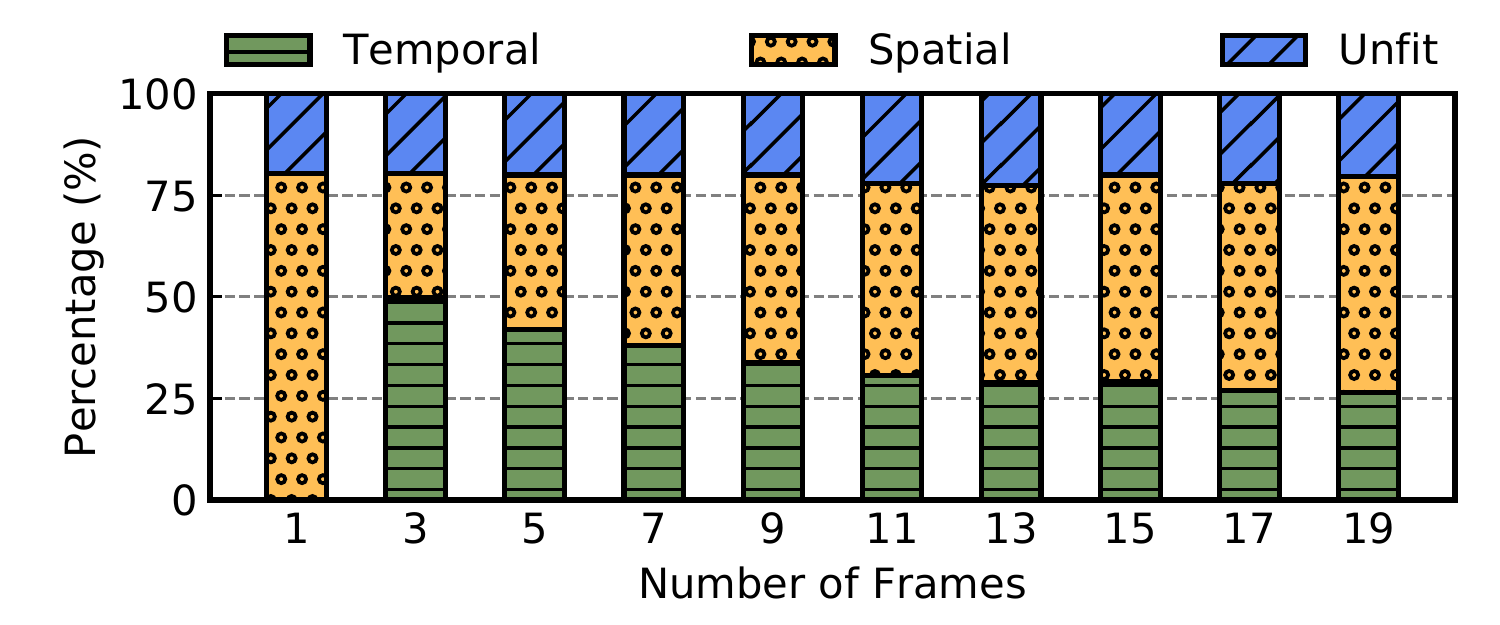}
\caption{Distribution of different encoding types within a point cloud sequence as the number of consecutively encoded frames varies. ``Unfit'' refers to points that could not be encoded in either method. Note that with only one frame there is no temporally encoded frame.}
\label{fig:channel_sensitivity}
%\vspace{-8pt}
\end{figure}

\begin{figure}[t]
%\vspace{-8pt}
\centering
\includegraphics[trim=0 0 0 0, clip, width=\columnwidth]{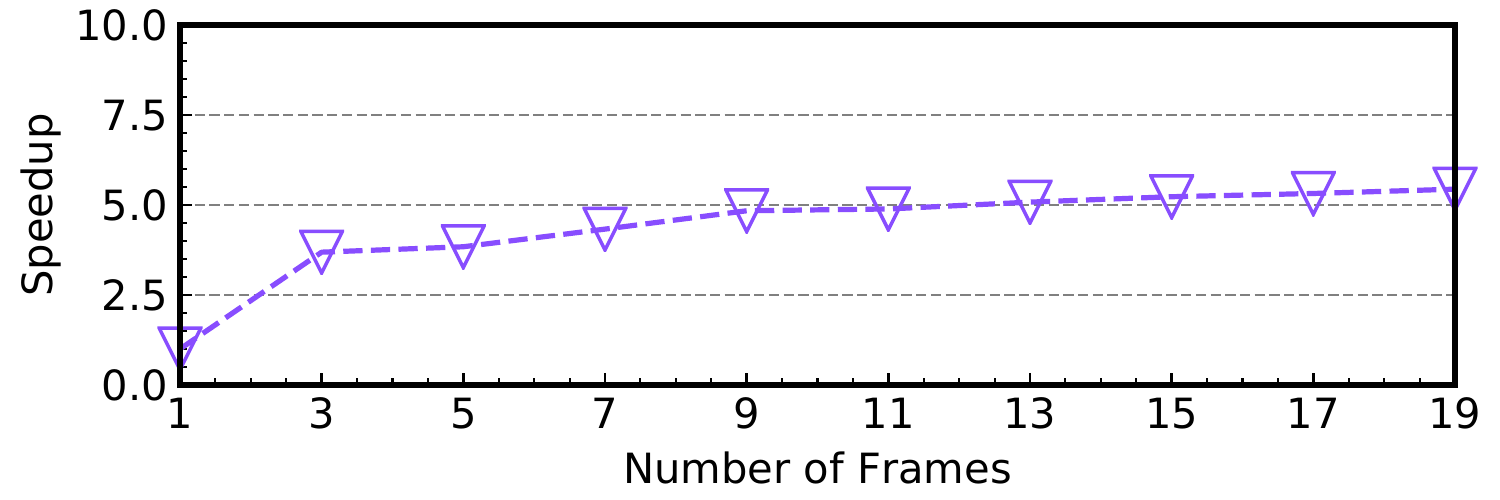}
\caption{Speedup of our parallel implementation over a sequential implementation on a four-core Intel i5-7500 CPU as the number of consecutive frames increases.}
\label{fig:channel_speedup}
%\vspace{-8pt}
\end{figure}

All results shown so far assume that five consecutive point clouds are encoded together. \Fig{fig:sensitivity} shows how compression rate, application accuracy, and compression rate vary with the number of consecutively encoded frames. All results are normalized to the results where the number of consecutive frames is five. Object detection uses individual frames, so its accuracy numbers are not shown.

We find that the compression speed is most sensitive to the number of encoded frames. This is because our implementation parallelizes many operations across frames such as the range image conversion and plane testing. More frames provides more opportunities for parallelization, leading to higher speeds. The application accuracies are mostly insensitive to the number of frames, because our compression method is able to preserve the vast majority of points during motion transformation and encoding.

It is worth noting that the compression rate is relatively insensitive to the number of frames. To understand why,~\Fig{fig:channel_sensitivity} shows the distribution of how different points are encoded in a sequence. As the the number of frames increases, the percentage of temporally encoded points decreases because the overlapped region becomes smaller, while the percentage of spatially encoded points increases. The overall percentage of points that are encoded by either methods stays roughly the same, leading to a roughly stable compression rate. Note that as the number of frames increases the decoding speed is faster with similar accuracy as shown in~\Fig{fig:sensitivity}, indicating longer sequences are preferred in encoding point clouds.

\Fig{fig:channel_speedup} shows the speedup of our parallel compression system over a sequential baseline. Recall from~\Sect{sec:med:opt} that our implementation exploits various forms of parallelism to improve the speed. With five frames available for compression, we achieve a 3.8$\times$ speedup over a sequential implementation. With 19 frames available, the speedup is 5.4$\times$. As the number of consecutive frames increases, the speedup saturates because of the hardware resource limitation.

\section{Future Work}
\label{sec:future}

While effectively, our existing compression algorithm also points out several areas of improvement for future work.

First, our current design uses native IMU measurements to register consecutive point cloud frames, which could suffer from inaccurate IMU measurements. Performing precise registration (e.g., using SLAM) could be computationally costly, defeating our goal of real-time compression. It would be interesting to investigate lightweight yet precise registration algorithms specifically designed for compression.

%, the precision of IMU measurement could potentially effect the compression performance. Instead, leveraging the real-time SLAM~\mbox{\cite{sun2017improving, sun2020see}} could be the potential direction to improve the transformation accuracy across multiple frames.

Second, our method extracts the geometric and temporal information from point cloud frames such as the plane coefficients, which could be used to assist certain robotic applications such as localization. Finally, it would be interesting to investigate point cloud algorithms that directly operate on compressed (encoded) point cloud.

\section{Conclusion}
\label{sec:conc}

Efficient point cloud compression will enable autonomous machines to be more connected with each other and with the cloud, and thus usher in a new era in distributed and cloud robotics. This paper proposes a novel spatio-temporal scheme for compressing LiDAR point clouds. We show that by exploiting spatial and temporal redundancies across consecutive point clouds, our compression method achieves up to 90$\times$ compression rate, maintains high application accuracy while achieving real-time ($>$10 FPS) compression speed. It out-performs the state-of-the-art point cloud compression standards on compression rate, speed, and accuracy.

%%%%%%%%%%%%%%%%%%%%%%%%%%%%%%%%%%%%%%%%%%%%%%%%%%%%%%%%%%%%%%%%%%%%%%%%%%%%%%%%

%%%%%%%%%%%%%%%%%%%%%%%%%%%%%%%%%%%%%%%%%%%%%%%%%%%%%%%%%%%%%%%%%%%%%%%%%%%%%%%%

%bibliography
\bibliographystyle{IEEEtranS}
\bibliography{refs}

\end{document}